# Structural Phase Transition and Carrier Density Tuning in SnSe$_x$Te$_{1-x}$ Nanoplates for Topological Crystalline Insulators

By *Jie Shen*[+], *John M. Woods*, *Y. Xie*, *M. D. Morales-Acosta,* and *Judy J. Cha**

[*]     Prof. J. J. Cha, Dr. J. Shen, J. M. Woods, Y. Xie
Department of Mechanical Engineering and Materials Science, Yale University, 15 Prospect St., New Haven, 06511 (USA)
Energy Sciences Institute, Yale West Campus, 810 West Campus Drive, West Haven, 06516 (USA)
Center for Research on Interface Structures and Phenomena, 15 Prospect, New Haven, 06511 (USA)
E-mail: judy.cha@yale.edu
        Dr. J. Shen
(present address) J.Shen-1@tudelft.nl
        M. D. Morales-Acosta
School of Applied Physics, Yale University, 15 Prospect St., New Haven, 06511 (USA)
Center for Research on Interface Structures and Phenomena, 15 Prospect, New Haven, 06511 (USA)

The major materials challenge facing topological insulators[1-3] is the high residual carrier density of the bulk,[4-6] which dominates over the surface state, making it difficult to observe transport properties of the surface state in nanoscale devices. Similar to topological insulators, the recently discovered topological crystalline insulators (TCIs)[7-8] such as SnTe suffer from the high bulk carrier density that masks the interesting surface states whose protection comes from the crystalline symmetry instead of the time reversal symmetry.[7-9] The high bulk hole carrier density of SnTe is due to Sn vacancies.[10-11] In order to study the surface states of TCIs, reducing the bulk carrier density is critical. One way to reduce the bulk carrier density is by making nanostructures, thereby largely removing the bulk.[12-14] Even in nanostructures however, the bulk carrier density is significant.[15] Recently, it was shown that the surface state can be revealed by deliberately making the bulk carrier mobility very low, which effectively freezes out the bulk contribution to the electrical transport at low temperature.[16-17] As a consequence, two-dimensional linear magnetoresistance that was sensitive to aging was observed in In-doped SnTe nanoplates.[17] Even in this case however, the residual bulk carrier density was still too high for any effective Fermi level tuning via electric field gating to expose only the surface states. Here, we show that by making ternary $SnSe_xTe_{1-x}$ nanostructures, we can decrease the bulk carrier density with increasing Se concentration. If resulting $SnSe_xTe_{1-x}$ nanostructures are cubic, they should be TCIs because cubic SnSe and SnTe have been experimentally shown to be TCIs.[18-23] However, as orthorhombic SnSe is more stable than cubic SnSe,[24] $SnSe_xTe_{1-x}$ can also be orthorhombic, thus not a TCI. We systematically map out the crystal structure evolution from the desired cubic structure to the unwanted orthorhombic structure as a function of the Se/Te ratio. In addition, we show how the temperature-dependent ferroelectric phase transition,[25-26] accompanied with the structural change from the high-temperature cubic to the low-temperature rhombohedral structure, evolves in cubic $SnSe_xTe_{1-x}$ nanoplates as a function of the Se/Te ratio. Our work represents a materials solution to solving the high bulk carrier density in TCIs and provides a pathway for exploring the surface states more effectively.

The exotic surface states of a TCI serve as a platform for fundamental scientific studies as well as future electronics such as low-dissipation electronics or spintronics. SnTe, a recently discovered TCI,[7-8] is a heavily doped p-type semiconductor. The bulk hole carriers effectively mask the electrical property of the surface states whose carrier density is significantly lower than the bulk carrier density. Recently, a cubic SnSe thin film was grown on $Bi_2Se_3$ by molecular beam epitaxy and shown to be a TCI and n-type semiconductor.[18] Therefore, we propose to synthesize ternary $SnSe_xTe_{1-x}$ nanoplates in which, at the right alloy composition of x, the carrier density of the nanoplate will be minimum, thus revealing the surface states. As long as the ternary structure is cubic, $SnSe_xTe_{1-x}$ should remain as a TCI because the topological protection comes from the crystalline symmetry.[7] The alloying strategy to reduce the bulk carrier density has been effectively demonstrated in the topological insulator alloy, $(Bi_xSb_{1-x})_2(Se_yTe_{1-y})_3$, where $Sb_2Te_3$ and $Sb_2Se_3$ are heavily p-type doped while $Bi_2Te_3$ and $Bi_2Se_3$ are heavily n-type doped.[27-28] Because $Bi_2Te_3$, $Bi_2Se_3$, and $Sb_2Te_3$ are topological insulators,[3] it was found that $(Bi_xSb_{1-x})_2(Se_yTe_{1-y})_3$ is also a topological insulator for a large range of x and y.[29] One caveat for the $SnSe_xTe_{1-x}$ alloy in our case is that a stable structure of SnSe is orthorhombic, not cubic.[18, 24] Thus, we must confirm that the $SnSe_xTe_{1-x}$ alloy nanostructure is cubic to ensure that it remains as a TCI.

We grow $SnSe_xTe_{1-x}$ alloy nanostructures by chemical vapor deposition using gold metal catalyst. A mixture of orthorhombic SnSe and cubic SnTe powders with a controlled molecular ratio of SnSe to SnTe is used as the source powder. $SiO_2$/Si substrates decorated with gold particles were used as growth substrates. Detailed growth procedure can be found in *Experimental Section*. The structure and chemical composition of the alloy nanostructures were characterized by transmission electron microscopy (TEM) and energy dispersive X-ray spectroscopy (EDX), as shown in **Figure 1**. Figure 1(a) – (f) correspond to a $SnSe_{0.2}Te_{0.8}$ nanoplate with $x_{nominal}$=0.2 indicating the nominal ratio. The EDX spectrum (Figure 1(g)) clearly shows co-presence of Se and Te in this nanostructure. We decompose the EDX spectrum into the reference spectra of SnSe and SnTe to calculate the real composition, $x_{EDX}$, in SnSe$_x$Te$_{1-x}$ (Supplementary Figure S1), which is measured to be 0.122 in this case (Figure 1(g)). The high angle annular dark field scanning TEM (HAADF STEM) image (Figure 1(a)) shows a 90 ° angle between two sides of the nanoplate (the red dashed lines), indicating that this is cubic. This is in agreement with the diffraction pattern (Figure 1(e)) and the high resolution TEM image (Figure 1(f)) that show cubic symmetry with 90 ° angles between diffraction spots and lattice fringes, respectively. X-ray diffraction (XRD) also confirms that the structure is cubic for SnSe$_{0.2}$Te$_{0.8}$ (**Figure 2**(a)). Elemental maps (Figure 1(b-d)) indicate a homogeneous distribution of Te and Se within the SnSe$_{0.2}$Te$_{0.8}$ nanoplate and show no segregation of either SnTe or SnSe. For the elemental mapping, we chose non-overlapping high energy peaks for Se (11.2 keV) and Te (27.4 keV).

When the nominal x is greater than 0.3, we find that the nanoplates are either single-crystalline cubic or single-crystalline orthorhombic. Figure 1(h)-(m) show an orthorhombic SnSe$_{0.5}$Te$_{0.5}$ (0.5 indicates the nominal ratio). Se and Te are again evenly distributed based on the elemental maps (Figure 1(i)-(k)). Unlike the cubic SnSe$_x$Te$_{1-x}$ however, the HAADF STEM image (Figure 1(h)) shows that the angle between the two sides of the plate is ~ 88 °, slightly less than 90 °. This is confirmed by electron diffraction (Figure 1(l)) and a high resolution TEM image (Figure 1(m)), in which the angles of the diffraction spots and lattice fringes deviate from the 90 ° red dashed lines. This is in agreement with orthorhombic SnSe, which has a facet angle of ~88°.[24, 30] The orthorhombic stucture of this nanoplate is due to the high Se concentration; the measured $x_{EDX}$ is 0.42, confirmed by the EDX fitting in Figure S1(c). We note that the Se/Te ratio, or the value of x, can be systematically controlled by tuning the mixture ratio of SnSe and

SnTe powders. Figure 1(o) shows the linear correlation between the nominal ratio $x_{nominal}$ we put in and the actual ratio $x_{EDX}$ we measure in corresponding SnSe$_x$Te$_{1-x}$ nanoplates by EDX. Each point is an average value from several nanoplates with the error bar indicating the standard deviation. Figure 1(n) shows respective EDX spectra of SnSe$_x$Te$_{1-x}$ with increasing $x_{nominal}$ of 0, 0.1, 0.2, 0.3, 0.4, 0.5, and 1 (shown in more detail in Figure S2). The decomposition analysis to calculate $x_{EDX}$ is shown in Supplementary Figure S1. We find the actual ratio is slightly less than the nominal ratio, likely due to different vapor pressures of SnSe and SnTe during growth.

With the increasing Se content, $x_{nominal}$, we find that the SnSe$_x$Te$_{1-x}$ structure changes from the desired cubic structure which ensures a TCI phase to the unwanted orthorhombic structure with the critical Se content, $x_c$, of ~ 0.3. Below $x_{nominal}$ of 0.3, all SnSe$_x$Te$_{1-x}$ nanoplates are cubic. Above $x_{nominal}$ of 0.3, the growth substrate containing SnSe$_x$Te$_{1-x}$ nanoplates shows a mixture of the cubic and orthorhombic phase. This was characterized by XRD. Figure 2 shows XRD patterns of SnSe$_x$Te$_{1-x}$ for $x_{nominal}$ = 0, 0.1, 0.2, 0.3, 0.4, 0.5, and 1. The XRD pattern from the pure SnSe (bottom XRD curve in dark-yellow in Figure 2(a)) matches the reference orthorhombic structure (Powder Diffraction File Card #00-048-1224, bottom dark-yellow bars) while the pure SnTe XRD pattern (top XRD curve in black) shows the reference cubic phase (Powder Diffraction File Card #00-046-1210, top dark bars). With x increasing up to 0.3, the XRD data obtained from the growth substrates show only the cubic phase, whose peaks are marked by the vertical dashed lines. The lattice constant systematically decreases with increasing x, as shown by the systematic shift in the (200), (220), (222), and (400) diffraction peaks of the cubic SnTe phase with reference to the vertical dashed lines. When $x_{nominal}$ is greater than 0.3, the XRD patterns from the growth substrates show mixture of the cubic and orthorhombic phases due to individual plates either being in the cubic or orthorhombic phase. The two phases can easily be distinguished from optical images by measuring the facet angle between two sides of the nanoplate, which is 90 ° for the cubic structure and slightly less than 90 ° for the orthorhombic structure (Figure 2(c) and (d)).

From the analysis of the XRD patterns, we extract lattice constants of the cubic phase of SnSe$_x$Te$_{1-x}$ as a function of x$_{nominal}$. Figure 2(b) shows the systematic decrease in the lattice constant of the cubic SnSe$_x$Te$_{1-x}$ with increasing x$_{nominal}$. This lattice constant was obtained from analyzing and averaging peak positions of (200), (220), (222) and (400). The decrease is expected as the cubic SnSe has a smaller lattice constant than the cubic SnTe. We note that the critical Se content x$_{nominal}$ at which the orthorhombic phase appears depends also on the growth substrate tempearature. For lower growth substrate temperature, the orthorhombic phase appears at lower x$_{nominal}$, indicating the orthorhombic structure is more stable (summarized in Supplementary Figure S3). Supplementary Figure S4 shows the phase diagram of SnSe$_x$Te$_{1-x}$ nanoplates on the growth substrates from the cubic to the orthorhombic phase as a function of the Se concentration x and substrate temperature.

We measure the carrier density of the SnSe$_x$Te$_{1-x}$ nanoplates as a function of x$_{nominal}$, as shown in **Figure 3** from Hall bar measurements. Typical Hall bar devices on cubic and orthorhombic nanoplates are on shown in Figure 2(c,d). Temperature-dependent resistance (R-T) and Hall data of pure SnTe and SnSe nanoplates are shown in Figure 3(a,b) and 3(k,l), respectively. As reported previously, pure SnTe nanoplates are heavily p-type doped[15] and the average carrier density is 3.7×10$^{21}$ cm$^{-3}$ at 1.8 K. In contrast, SnSe nanoplates exhibit semiconducting behavior as evidenced by the increasing R at lower T (Figure 3(k)). The Hall measurements of four SnSe devices show the average carrier density of SnSe is $2.7 \times 10^{17}$ cm$^{-3}$ at room temprature and decreases dramatically with decreasing temperature. Interestingly, our orthorhombic SnSe nanostructures are p-type while the MBE grown cubic SnSe thin film on Bi$_2$Se$_3$ was n-type.[18] With increasing x in SnSe$_x$Te$_{1-x}$, the Hall slope at 1.8 K systematically increases from 0.35 Ω change over 9 T for SnSe$_{0.1}$Te$_{0.9}$ to 0.57 Ω, 78 Ω, and 117 Ω changes over 9 T for x$_{nominal}$ = 0.2, 0.4, and 0.5 respectively. The corresponding carrier densities are $1.6 \times 10^{21}$ cm$^{-3}$, $9.1 \times 10^{20}$ cm$^{-3}$, $2.9 \times 10^{18}$ cm$^{-3}$, and $1.9 \times 10^{18}$ cm$^{-3}$, clearly showing the systematic decrease in the carrier density with increasing x. Summary of the transport data as a function of x is presented in **Table 1** and Figure S5. The systematic control of the carrier density change observed in SnSe$_x$Te$_{1-x}$ is similar to the results for Bi$_2$(Se$_x$Te$_{1-x}$)$_3$ and (Sb$_x$Bi$_{1-x}$)$_2$Te$_3$ nanoplates.[28, 31] In contrast to the systematic change of the carrier density with x, mobility decreases significantly for the ternary nanoplates compared to SnTe (Table 1 and Supplementary Figure S5).

The structural transition from the cubic to the orthorhombic phase in SnSe$_x$Te$_{1-x}$ at x=0.3 can also be observed in the electrical tranport data. We note a kink, marked by purple arrows, is observed in the R-T curves in Figure 3 (a, c, e) when x < 0.3 (Device #1, #2, #3), which indicates the structural phase transition from the cubic to rhombohedral phase at low temperature, in agreement with previous reports for cubic SnTe.[32-33] In contrast, for x > 0.3 (Device #4, #5, #6), the R-T behavior shows no kinks and increases at low temperatures, suggesting that the Fermi level must be close to the band edge or in the gap (Figure 3(g, i, k)). The continuous increase in R without saturation at low temperatures may indicate the absence of the surface states, which are not expected in the orthorhombic nanoplates. The purple dashed line between Figure 3(e) and 3(g) is drawn to show the transition of the two different structures.

For SnTe, a structural phase transition from the high temperature cubic to the low temperature rhombohedral structure is known to occur, which is accompanied with a ferroelectric phase transition.[33-34] This has been observed in the bulk crystal study as well as in the SnTe nanoplate study.[15] The structural phase transition is manifested as a kink in the temperature-dependent resistivity. In the $SnSe_xTe_{1-x}$ cubic nanoplates, we observe a systematic trend in the phase transition as a function of $x_{nominal}$, as shown in **Figure 4**. Figure 4(a) shows temperature-dependent resistance normalized to the resistance at 1.8K for Device #1 (x=0), Device #2, #2a, #2b ($x_{nominal}$=0.1), and Device #3, #3a, #3b, #3c ($x_{nominal}$=0.2). For pure SnTe (Deivce #1), we see a clear kink, indicated by the arrow. For $SnSe_xTe_{1-x}$ however, instead of a sharp kink at a singular temperature value, we observe a broad peak in resistance, which spans over tens of degrees in temperature. To highlight the differences, we plot dR/dT, shown in Figure 4b (more details in Figure S6 (a)). We make a number of observations. First, we see a peak around 48 K ($T_{down}$) in dR/dT curves that are independent of the Se content, $x_{nominal}$. The peak broadens for higher $x_{nominal}$, indicated by the blue shaded boxes that denote the starting point of the transition (Tup). The width of the peak, $\Delta T=Tup-Tdown$, increases with increasing xnominal. These parameters are summarized in the bottom part of Table1. We denote Tup as the starting point of the structural transition from the cubic phase to the rhombahedral phase and Tdown as the ending point. The ferroelectric phase transition of the cubic SnTe is related to the carrier density according to the soft transverse optical phonon model.[33, 35-36] We plotted Tup and $\Delta T$ as a function of xnominal and carrier density n in Figure 4(c) and (d) respectively and found that both Tup and ΔT have a linear relation with xnominal rather than n. This results show that the compositional change is the dominant effect, similar to phase transition temperatures in $Pb_{1-x}Ge_xTe$,[25, 37] $Ge_{1-x}Sn_xTe$,[26, 38] and $Pb_{1-x}Sn_xTe$[39] ternary compounds. Because the cubic SnSe has a higher phase transition temperature than the cubic SnTe,[40] the transition temperature of $SnSe_xTe_{1-x}$ is between that of SnSe and SnTe, and increases with the increasing Se content. For $SnSe_xTe_{1-x}$, Figure 4(c) represents the first phase diagram from the cubic to the rhombahedral structure. We attribute the increasing width of the structural transition, ΔT, to the increasing disorder[41] in $SnSe_xTe_{1-x}$ with xnominal although the exact microstructural origin is unclear at present. More details about the phase diagram are shown in Figure S6(b).

We have demonstrated that the carrier density of the ternary $SnSe_xTe_{1-x}$ nanoplates can be systematically decreased with increasing Se content. This material alloying strategy presents an effective solution to lower the bulk residual carrier density which often masks the surface states in TCIs. We also found that with increasing Se content, the growth products, $SnSe_xTe_{1-x}$ nanoplates, can either be cubic or orthorhombic. When xnominal < 0.3, all $SnSe_xTe_{1-x}$ nanoplates are cubic, which ensures that they are TCIs. When xnominal >0.3, $SnSe_xTe_{1-x}$ nanoplates can be either cubic or orthorhombic. Distinguishing the two structures is straightforward by measuring the facet angles between the two sides of the nanoplates in optical images, which is 90 ° for the cubic structure and less than 90 ° for the orthorhombic structure. The cubic to rhombohedral structural transition is also observed for cubic $SnSe_xTe_{1-x}$ where the starting point of the structural transition increases with the increasing x. This is consistent with previous observations in $Pb_{1-x}Ge_xTe$, $Ge_{1-x}Sn_xTe$, and $Pb_{1-x}Sn_xTe$ alloys. Moreover, we observe that the structural transition occurs gradually for the alloy, which is attributed to the increasing disorder in the alloy system in which the Se and Te are randomly occupying the sites. These distinct electrical properties in the SnSexTe1-x alloy nanoplates can provide a new strategy to realize electronic and spintronic devices based on the TCI.

*Experimental Section*

*Synthesis of SnSexTe1-x:* SnSexTe1-x nanoplates were synthesized using a tube furnace with a 1 inch horizontal tube. SiOx/Si substrates decorated with Au particles were used as growth substrates. A mixture of SnSe and SnTe powders was placed at the center of the furnace at 600 °C while the growth substrates were placed at downstream in the temperature range of 300–510 °C(Supplementary Figure S3). The temperatures for the four growth substrates were 510 °C, 440 °C, 370 °C, and 300 °C, respectively. Argon gas was used as the carrier gas with the flow rate of 80 sccm and the pressure of ~1 Torr during growth. The furnace was ramped to the growth temperature in 30 minutes and kept at the growth temperature for 10 minutes, before the natural cool down. Visual inspection with an optical microscope was sufficient to distinguish the cubic nanoplates from the orthorhombic nanoplates.

*Characterizations:* The synthesized nanoplates were characterized structurally using transmission electron microscopy (TEM, FEI Tecnai Osiris 200 keV) and X-ray diffraction (Rigaku SmartLab Diffractometer with Cu Kα radiation ($\lambda=1.54$Å) at the Center for Research on Interface Structures and Phenomena (CRISP)). Electron diffraction patterns and high-resolution TEM images were obtained to analyze the cubic and the orthorhombic phase. Energy-dispersive

X-ray spectroscopy (EDX) equipped in the TEM was used to calculate the composition of $SnSe_xTe_{1-x}$.

*Device Fabrication and Transport Measurement*: Nanoscale devices on $SnSe_xTe_{1-x}$ nanoplates were fabricated following the standard e-beam lithography (Vistec EBPG 5000+).  10nm/200nm Cr/Au metal contacts were deposited by thermal evaporation (MBraun MB-EcoVap 4G).  Before metallization, the contact areas were etched with Ar plasma for Ohmic contact. Physical Property Measurement System (Quantum Designs, Dynacool) was used to measure the electrical transport properties of $SnSe_xTe_{1-x}$ nanodevices in temperatures down to 1.8 K and magnetic fields up to 9 T. All the Hall curves were measured at 1.8 K except for pure SnSe, which was measured at room temperature.

## *Supporting Information*

Supporting Information is available from the Wiley Online Library or from the author.

## *Acknowledgements*

This work was supported by Department of Energy Basic Energy Science, Award No. DE-SC0014476.  J. S. was partly supported by NSF DMR 1402600.  Microscopy facilities used in this work were supported by the Yale Institute for Nanoscience and Quantum Engineering (YINQE) and MRSEC DMR 1119826.

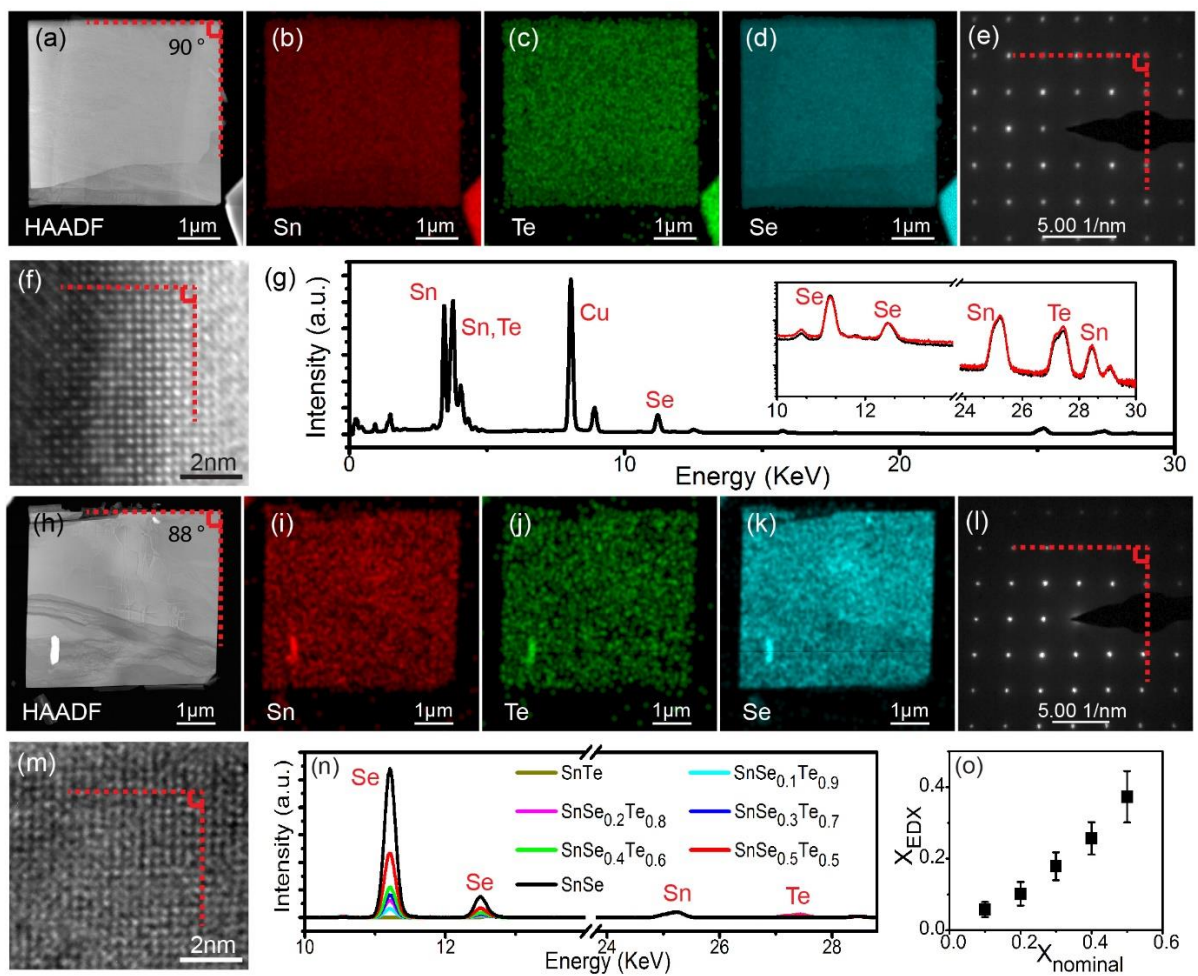

**Figure 1.** Structural characterization of SnSe$_x$Te$_{1-x}$ nanoplates for $0 \leq x \leq 1$. (a) High-angle annular dark field STEM image of SnSe$_{0.2}$Te$_{0.8}$. $x_{nominal}$ of 0.2 represents the nominal ratio. Chemical maps of (b) Sn, (c) Te, and (d) Se, which show uniform distribution of all elements. (e) Diffraction pattern of the SnSe$_{0.2}$Te$_{0.8}$ nanoplate shows cubic symmetry with a 90° angle between the spots, confirming the cubic phase. (f) High-resolution TEM image that shows lattice fringes with a 90° angle between two fringe directions. (g) EDX spectrum used to calculate the Se/Te ratio. The red curve in the inset shows the fitting result of $x_{EDX}$=0.122, which gives good agreement with the data (black curve). The presence of Cu is from the Cu TEM grid. (h) High-angle annular dark field STEM image of SnSe$_{0.5}$Te$_{0.5}$. Chemical maps of uniformly distributed (i) Sn, (j) Te, and (k) Se. (l) Diffraction pattern of the SnSe$_{0.5}$Te$_{0.5}$ shows off-cubic

symmetry where the angle between the two primary axes of the diffraction spots is slightly less than 90°. (m) High-resolution TEM image shows that the two lattice fringe directions are not at 90°, but less than 90°, indicating the orthorhombic phase. (n) EDX spectra of SnSe$_x$Te$_{1-x}$ with varying x. The Se peak systematically increases while the Te peak decreases with increasing x. (o) Linear correlation between the nominal ratio x$_{nominal}$ and the measured ratio x$_{EDX}$ in the SnSe$_x$Te$_{1-x}$ nanoplates. All the red dashed lines show 90° angles.

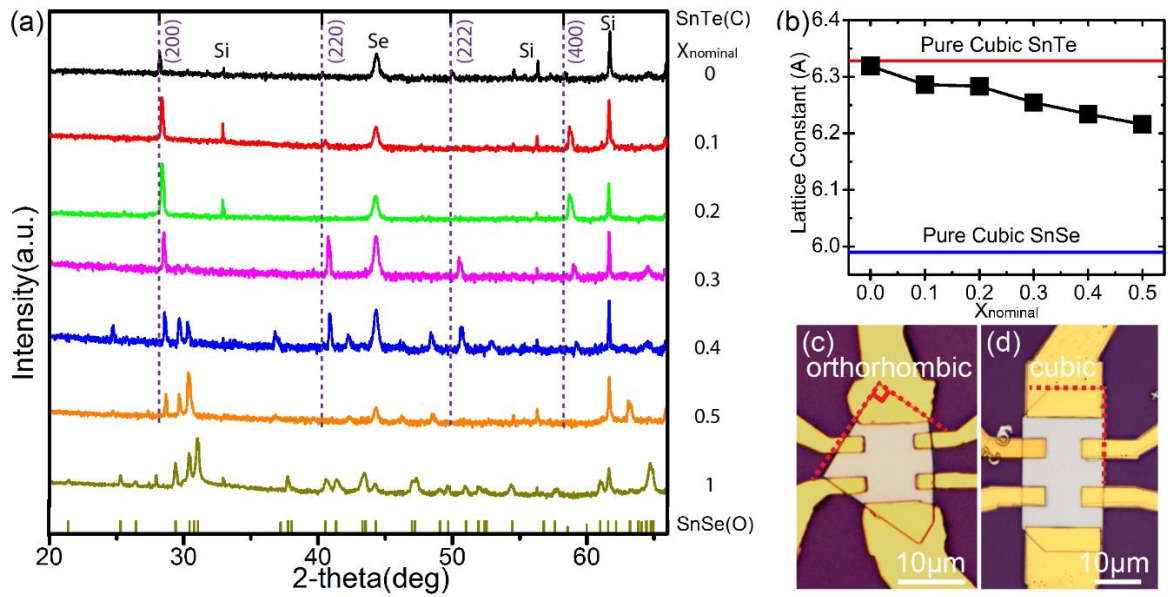

**Figure 2.** X-Ray Diffraction of SnSe$_x$Te$_{1-x}$ for $0 \leq x \leq 1$. (a) X-ray diffraction patterns for x$_{nominal}$ = {0, 0.1, 0.2, 0.3, 0.4, 0.5 & 1}. Reference peak positions for cubic SnTe and orthorhombic SnSe are shown at the top and bottom axes, respectively. Vertical dashed lines indicate (200), (220), (222), and (400) planes of cubic SnTe. (b) Systematic decrease in the lattice constant as a function of the Se concentration x$_{nominal}$ for the cubic phase SnSe$_x$Te$_{1-x}$. The lattice constants for the cubic SnTe (red) and the cubic SnSe (blue) are shown for comparison. (c) Orthorhombic SnSe$_x$Te$_{1-x}$ nanoplate which shows a facet angle less than 90° between two sides, marked by the red dotted lines. (d) Cubic SnSe$_x$Te$_{1-x}$ nanoplate which shows a 90° facet angle between two sides, marked by the red dashed lines. These two nanoplates have been made into hall bars, with the Cr/Au contacts.

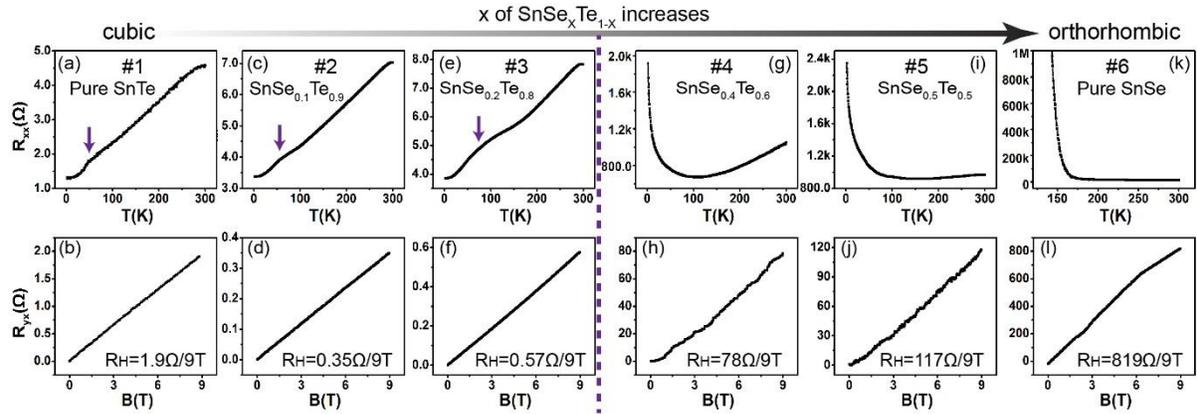

**Figure 3.** Transport properties of SnSe$_x$Te$_{1-x}$ for $0 \leq x \leq 1$. Temperature-dependent resistance and Hall resistance of cubic SnTe Device #1 (a,b); SnSe$_{0.1}$Te$_{0.9}$ Device #2 (c,d); SnSe$_{0.2}$Te$_{0.8}$ Device #3 (e,f); and orthorhombic SnSe$_{0.4}$Te$_{0.6}$ Device #4 (g,h); SnSe$_{0.5}$Te$_{0.5}$ Device #5 (i,j); SnSe Device #6 (k,l). For $x_{nominal} < 0.3$, the resistance decreases with decreasing temperature and shows a kink that is related to the structural transition to the rhombohedral phase. For $x_{nominal} > 0.3$, the resistance increases with decreasing temperature, indicating either the Fermi level is close to the band edge or in the bulk gap. The Hall slope systematically increases with increasing x, indicating decrease in the carrier density with increasing x. The purple arrows in (a), (c) and (e) show the phase transition from the cubic to rhombohedral phase at low temperature. All the Hall curves were obtained at 1.8 K, except for SnSe shown in (l), which was measured at room temperature. Transport data of Device#1 is from Ref [15].

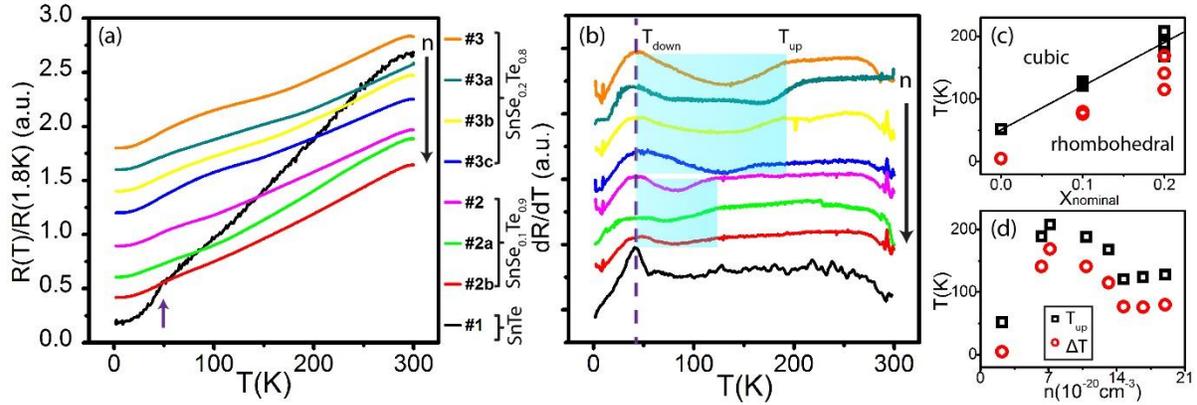

**Figure 4.** Structural transition from the cubic phase to the rhombohedral phase with temperature for cubic SnSe$_x$Te$_{1-x}$ nanoplates (Device #1 - #3). (a) Normalized resistance curves as a function of temperature show a kink for SnTe and broad peaks for SnSe$_x$Te$_{1-x}$, which indicates structural transition from the high-temperature cubic phase to the low-temperature rhombohedral phase. The normalized resistance curves were plotted with decreasing Se concentration $x_{nominal}$ (thus increasing carrier density) from top to bottom. (b) *dR/dT* of the same devices shown in panel (a). The structural transition appears to end at the same temperature of ~ 48 K, indicated by the vertical dashed line. The start of the structural transition depends on the Se concentration, $x_{nominal}$, as indicated by the shaded cyan boxes. The *dR/dT* curves were plotted with decreasing Se concentration $x_{nominal}$ (thus increasing carrier density) from top to bottom. (c) and (d) show the starting point ($T_{up}$, black squares) and the width of the transition ($\Delta T$, red dots) versus the Se concentration $x_{nominal}$ and the carrier density n, respectively. The linear relation between the transition temperature and the composition ratio, $x_{nominal}$, shown in (c) agrees with the previous reports on GeTe-SnTe and PbTe-SnTe materials. Peak broadening observed in SnSe$_x$Te$_{1-x}$ is attributed to the disorder, although the detailed mechanism is unknown. No clear linear relations can be observed between the transition temperature and the carrier density, as shown in (d), suggesting that the carrier density may not affect the transition significantly.

**Table 1.** Summary of electrical transport properties of SnSe$_x$Te$_{1-x}$ nanoplates for $0 \leq x \leq 1$. The parameters of SnSe devices, Device #6, are measured at room temperature because Fermi level in these devices is close to the band gap and the devices are insulating at 1.8K. The top part of the table shows the average carrier density ($\bar{n}$) and the average mobility ($\bar{\mu}$) from several devices with the same alloy ratio. The bottom part of the table contains all the parameters of devices shown in Figure 4.

| X$_{nominal}$ | 0 | 0.1 | 0.2 | 0.4 | 0.5 | 1 |
|---|---|---|---|---|---|---|
| Device | #1, … | #2, 2a, 2b | #3, 3a, 3b, 3c | #4, … | #5, … | #6, … |
| structure | cubic | cubic | cubic | orthorhombic | orthorhombic | orthorhombic |
| $\bar{n}$(cm$^{-3}$) | $(3.7\pm3.5)\times10^{21}$ | $(1.6\pm0.2)\times10^{21}$ | $(9.1\pm3.1)\times10^{20}$ | $(2.9\pm1.1)\times10^{18}$ | $(1.9\pm0.8)\times10^{18}$ | $(2.7\pm2.1)\times10^{17}$ |
| $\bar{\mu}$(cm$^2$/Vs) | 156.7±67.3 | 30.1±4.7 | 49.2±9.3 | 11.3±4.5 | 13.5±5.2 | 22.7±7.2 |

| Device | #1 | #2b | #2a | #2 | #3c | #3b | #3a | #3 |
|---|---|---|---|---|---|---|---|---|
| n (cm$^{-3}$) | $2.2\times10^{20}$ | $18.9\times10^{20}$ | $16.7\times10^{20}$ | $14.7\times10^{20}$ | $13.1\times10^{20}$ | $10.8\times10^{20}$ | $7.1\times10^{20}$ | $6.2\times10^{20}$ |
| μ (cm$^2$/Vs) | 670 | 34.3 | 30.7 | 24.9 | 35.1 | 34.9 | 75.9 | 50.9 |
| Thickness(nm) | 109 | 182 | 131 | 110 | 132 | 128 | 154 | 156 |
| T$_{down}$(K) | 47 | 48 | 48 | 46 | 53 | 47 | 39 | 48 |
| T$_{up}$(K) | 52 | 128 | 124 | 121 | 168 | 188 | 208 | 189 |
| ΔT(K) | 5 | 80 | 76 | 77 | 115 | 141 | 169 | 141 |

**Cubic SnSe$_x$Te$_{1-x}$** is expected to be a topological crystalline insulator whose surface states are protected by crystal symmetry. SnSe$_x$Te$_{1-x}$ nanoplates with $0 \leq x \leq 1$ are synthesized in the cubic and orthorhombic phase to show systematic decrease of the bulk carrier density with increasing x. Ferroelectric phase transition is also observed in these nanoplates.

Keyword: Topological Crystalline Insulator, SnTe, SnSe$_x$Te$_{1-x}$, Electronic Transport, Structural Phase Transition

J. Shen, J. M. Woods, Y. Xie, M. D. Morales-Acosta, J. J. Cha*

**Structural Phase Transition and Carrier Density Tuning in SnSe$_x$Te$_{1-x}$ Nanoplates for Topological Crystalline Insulators**

ToC figure

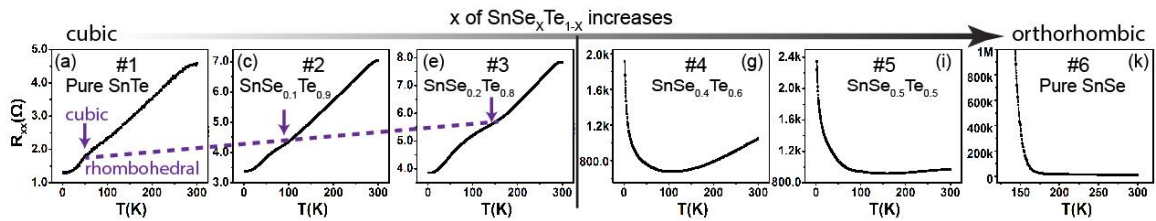

**Supporting Information**

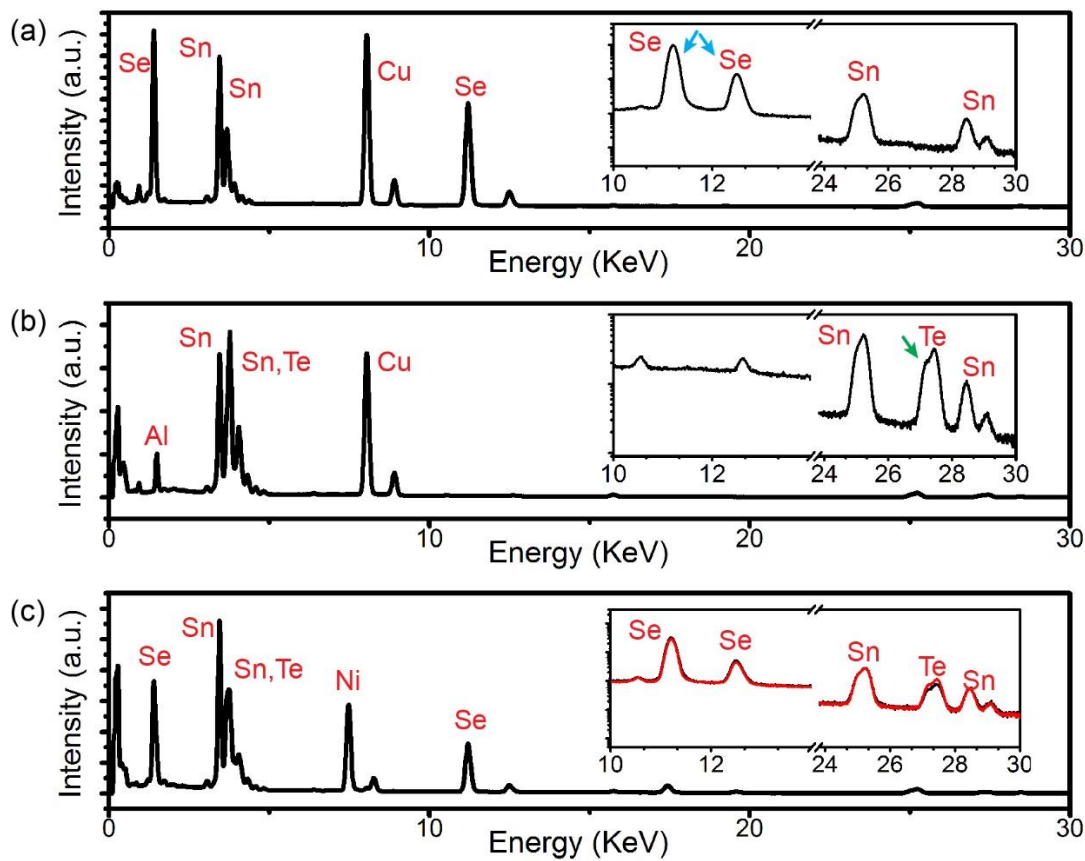

**Supplementary Figure S1.** Reference EDX spectra of SnSe (a) and SnTe (b). The blue and green arrows in the insets point to the Se and Te peaks respectively. (c) shows the EDX spectrum of the SnSe$_{0.5}$Te$_{0.5}$ ($x_{nominal}$=0.5), shown in Figure 1(h-m). The red line is a fit to the data, which gives $x_{EDX}$ of 0.422. The value of $x_{EDX}$ was obtained by fitting the spectrum with the linear combination of the SnTe and SnSe reference spectra. Cu and Ni in the EDX spectra are from the Cu and Ni TEM grids.

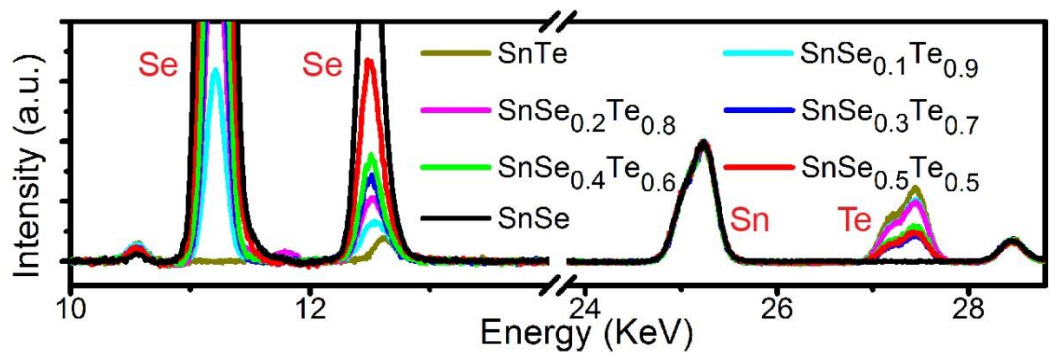

**Supplementary Figure S2**. EDX spectra of SnSe$_x$Te$_{1-x}$ for increasing x. The Se peak systematically increases while the Te peak decreases with increasing x.

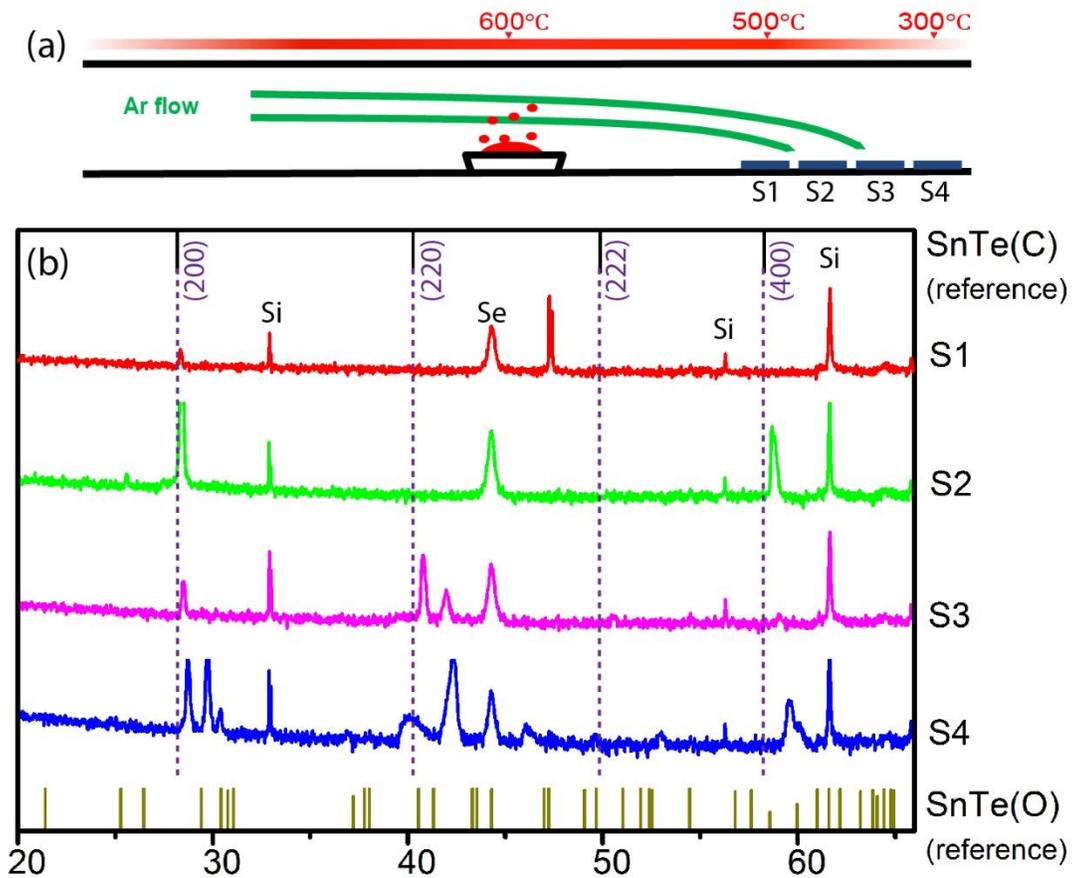

**Supplementary Figure S3**. (a) Growth schematic figure. (b) X-ray diffraction patterns of $SnSe_xTe_{1-x}$ of different substrate temperatures for the same nominal ratio of 0.2. The substrate temperature also influences the structure even for the same value of $x_{nominal}$. Substrate 1 (S1) was placed at the highest temperature of 510 °C while Substrate 4 (S4) was placed at the lowest temperature of 300 °C. The orthorhombic phase is more prominent at lower temperature substrates (S4).

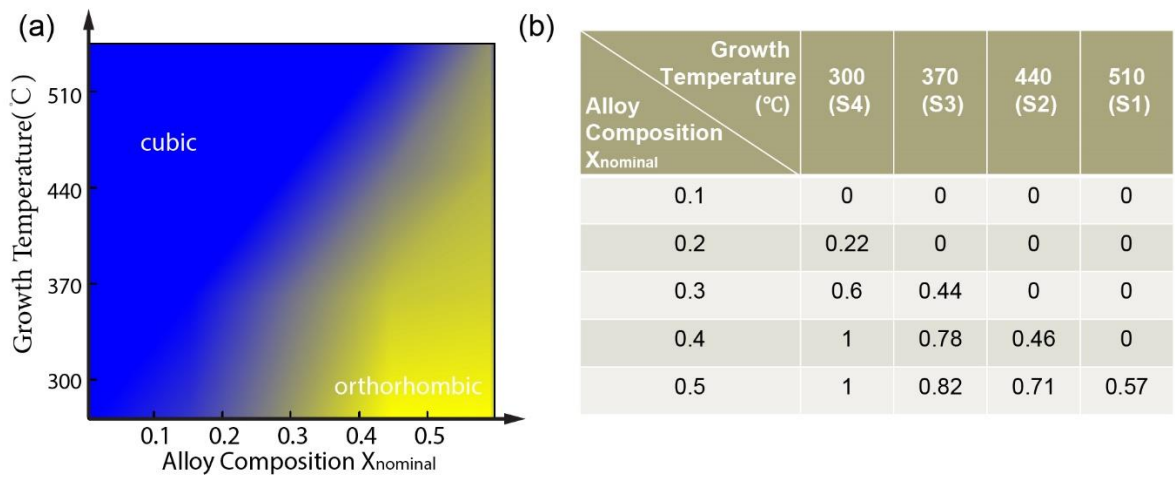

**Supplementary Figure S4**. (a) Schematic of the cubic versus orthorhombic phase diagram as a function of the Se concentration in SnSe$_x$Te$_{1-x}$ and the substrate temperature. (b) A summary table of the phase diagram. Numbers indicate the ratio of the cubic peak at 28.2° to the orthorhombic peak at 30.7° from XRD data. "0" indicates only the cubic structure while "1" indicates only the orthorhombic structure. The phase diagram shown in (a) is plotted based on these numbers.

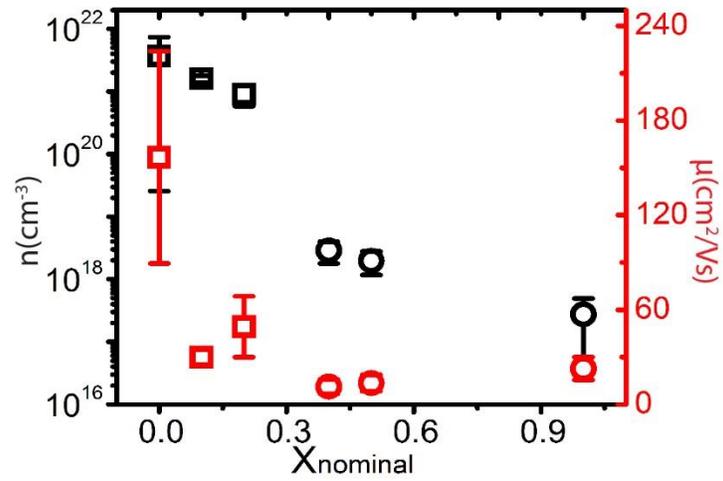

**Supplementary Figure S5**. Carrier density and mobility of SnSe$_x$Te$_{1-x}$ nanoplates for $0 \leq x \leq 1$. The carrier density (black) decreases systematically for increasing Se concentration, $x_{nominal}$. The mobility (red) drops dramatically for the alloy nanoplates. Squares indicate the cubic structure while the circles indicate the orthorhombic structure.

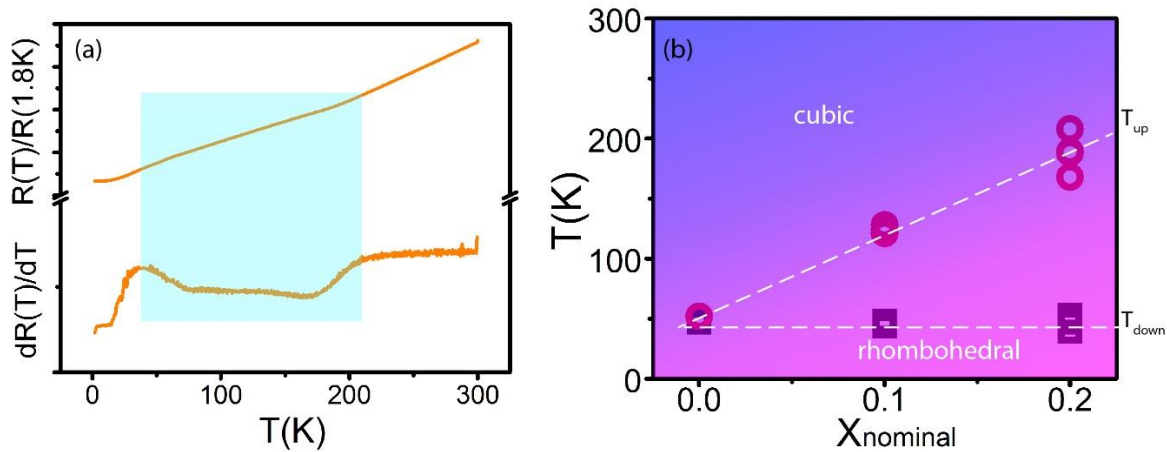

**Supplementary Figure S6**. (a) shows the R/R(1.8 K) and dR/dT of Device #3 as a function of temperature, which reflects the gradual transition from the cubic to the rhombohedral phase of Device #3, SnSe$_{0.2}$Te$_{0.8}$. (b) shows the phase diagram from the cubic to the rhombohedral phase with temperature (y-axis) and alloying composition, x$_{nominal}$ (x-axis). Blue represents the cubic structure while purple represents the rhombohedral structure.